\documentclass[12 pt]{iopart}

\usepackage{iopams}
\usepackage{graphicx} 
\usepackage[format=plain,justification=justified,singlelinecheck=false,font={stretch=1.125,small,sf},labelfont=bf,labelsep=colon]{caption}
\usepackage{subcaption}
\usepackage{array}
\usepackage{balance}
\usepackage[dvipsnames]{xcolor}
\usepackage{hyperref}
\hypersetup{colorlinks=true, linkcolor=blue, citecolor=blue, urlcolor=blue}
\usepackage{setspace}

\usepackage{graphicx}   % For including graphics
\usepackage{subcaption} % For subfigure captions

\usepackage{ulem}
\usepackage{bm} 
\usepackage{nicefrac}
\usepackage{xcolor}

\usepackage[separate-uncertainty=true]{siunitx}
\DeclareSIUnit\angstrom{\textup{~\AA}}
\usepackage{xspace}
\newcommand{\ie}{\textit{i.\,e.}~}

\newcommand{\pfunit}[1]{#1Wm$^{-1}$K$^{-2}$}
\newcommand{\klunit}[1]{#1Wm$^{-1}$K$^{-1}$}

\begin{document}

\title[PbXY Janus layer]{Thermoelectric properties of Lead halide Janus layers - A theoretical investigation}
\author{A.E. Sudheer} \address{Department of Sciences, Indian Institute of Information Technology Design and Manufacturing Kurnool Andhra Pradesh, 518007, India}
\author{M Vallinayagam} \address{Institute of Experimental Physics, TU Bergakademie Freiberg, Leipziger Str. 23, 09596 Freiberg, Germany}\address{Technical Physics, University of Applied Sciences, Friedrich-List-Platz 1, 01069 Dresden, Germany}
\author{G Tejaswini} \address{Department of Sciences, Indian Institute of Information Technology Design and Manufacturing Kurnool Andhra Pradesh, 518007, India}
\author{Amrendra Kumar} \address{Theory and Simulations Laboratory, Theoretical and Computational Physics Section, Raja Ramanna Centre for Advanced Technology, Indore 452013, India} \address{Homi Bhabha National Institute, Training School Complex, Mumbai 400094, India}
\author{M Posselt} \address{Helmholtz-Zentrum Dresden-Rossendorf, Institute of Ion Beam Physics and Materials Research, 01328 Dresden, Germany}
\author{C Kamal} \address{Theory and Simulations Laboratory, Theoretical and Computational Physics Section, Raja Ramanna Centre for Advanced Technology, Indore 452013, India} \address{Homi Bhabha National Institute, Training School Complex, Mumbai 400094, India}
\author{M Zschornak} \address{Institute of Experimental Physics, TU Bergakademie Freiberg, Leipziger Str. 23, 09596 Freiberg, Germany} \address{Technical Physics, University of Applied Sciences, Friedrich-List-Platz 1, 01069 Dresden, Germany} \ead{matthias.zschornak@htw-dresden.de}
\author{D Murali} \address{Department of Sciences, Indian Institute of Information Technology Design and Manufacturing Kurnool Andhra Pradesh, 518007, India} \ead{dmurali@iiitk.ac.in}

\vspace{10pt}
\begin{indented}
\item[]\today
\end{indented}

\vspace{10pt}

\begin{abstract}
Thermoelectric materials offer a promising route for efficient heat-to-power conversion. In search of materials functional at high and low operating temperatures, we investigate the thermoelectric properties of two-dimensional lead halide Janus layers (JLs) using density functional theory. The electronegativity difference between halides in JLs significantly modulates the electronic structure, particularly the strong Pb–F bonding in PbIF JLs leads to pronounced band curvature and a unique direct bandgap. Estimated through three-phonon interactions, the lattice thermal conductivity is intrinsically low, primarily due to acoustic phonon contributions and suppressed optical phonon transport. The thermoelectric coefficients are enhanced with carrier doping, resulting in figures of merit as high as 8.94 at room temperature and upto 36.31 at elevated temperatures. These findings establish two-dimensional lead halide Janus layers as exceptional candidates for thermoelectric conversion, and the insights into their elemental and electronic characteristics offer a valuable basis for the future design of high-performance lead-based thermoelectric materials.
\end{abstract}

% \submitto{\TDM}
% % Uncomment if a separate title page is required
% \maketitle
% \ioptwocol

\section{\label{intro}Introduction}
The growing concern over global climate change, driven by fossil fuel consumption, underscores the urgent need for green energy technologies. Among these, thermoelectric (TE) materials offer a promising solution by directly converting temperature gradients into electricity via the Seebeck effect. This effect establishes a voltage, $V=S\cdot\Delta T$, where $S$ is Seebeck coefficient~\cite{snyder08_natmat}. The efficiency of TE materials is quantified by the dimensionless parameter figure of merit $ZT$, defined as~\cite{snyder08_natmat, Gorai2017, Shu2024, Xiao2018_natmat, Cao2023_eSci, binbin2021_sci, Mamani2024}:
\begin{equation}
    ZT~=~S^2 \sigma T/\kappa, \label{eq:ZT}
\end{equation}
where $\sigma$ and $\kappa$ are electrical and total thermal conductivity, respectively. The total thermal conductivity comprises two components: electronic $\kappa_e$, arising from free carrier movement, and lattice $\kappa_l$, due to atomic vibrations. Optimizing $ZT$ is challenging due to the interdependent nature of key TE parameters ($S$, $\sigma$, and $\kappa$). These parameters depend on charge carrier concentration $N$, following the trends $S \propto N^{-2/3}$, $\sigma \propto N$, and $\kappa_e \propto N$~\cite{snyder08_natmat}. While increasing $N$ enhances $\sigma$, it simultaneously reduces $S$. Since $ZT$ is proportional to the power factor $PF=S^2 \sigma$ and inversely proportional to $\kappa$, achieving high TE performance requires a balancing act, maximizing both $S$ and $\sigma$ while minimizing $\kappa$~\cite{Xiao2018_natmat, Bai2023}.

Various strategies have been explored to achieve high $ZT$, including band engineering~\cite{Pei2011, Park2021, Hu2021, Zhu2022, Zhang2025}, strain engineering~\cite{wu2022, Shulin2022, Chang2025, Xoiong2025}, entropy engineering~\cite{Jiang2021, Schweidler2024}, structural modifications~\cite{Bjing2021, Qin2023, Wang2024, Cwu2024, Jdong2024}, doping~\cite{qin2022ami, Dadhich2023, Qxin2024}, alloy formation~\cite{kumar2024jpcc, binbin2021}, and symmetry breaking~\cite{murari2024, Ould2024}. For example, the Bi doping in GeTe is shown to reduce the majority carrier concentration, enhancing the carrier mobility, which results in a power factor of 28.40 \pfunit{$\mu$} and a maximum $ZT$ of 2.1 at $\approx$ 600 K~\cite{Angran2025}. Additionally, the strength of spin-orbit interactions has been shown to influence the thermoelectric power factor, with materials exhibiting a Rashba parameter $\alpha_R$ in the range of 2.75–3.55~\si{\electronvolt\angstrom} demonstrating optimal performance~\cite{Li2020ncm}. Given the intricate interdependence of TE parameters, recent theoretical efforts have focused on identifying materials with intrinsically low thermal conductivity $\kappa$. In particular, optimizing phonon and electronic transport by minimizing the phonon mean free path has emerged as a key strategy for maximizing $ZT$~\cite{AEIVARI2021, CDing2023}.

Among these strategies, structural modifications have attracted significant attention for their effectiveness in reducing $\kappa_l$, enabling ultralow thermal conductivity. For instance, two-dimensional (2D) allotropes of Si and Ge exhibit substantially lower $\kappa_l$ values of 17.08~\klunit{} and 1.10~\klunit{}, respectively, compared to their bulk counterparts~\cite{Shu2024}. The Pb$_2$XY monolayer achieves a remarkable $ZT$ of 6.88 at 800 K, attributed to its low $\kappa_l$ of 0.162~\klunit{}, driven by strong phonon scattering and reduced group velocity~\cite{Mamani2024}. Similarly, in hydrogenated 2D layers Si$_2$SbH$_2$ and Ge$_2$SbH$_2$, buckling-induced flexural phonons significantly suppress $\kappa_l$ to 0.12~\klunit{}~\cite{Mohebpour2021}. Additionally, point defects such as vacancies and atomic off-centering further reduce $\kappa_l$ by enhancing phonon scattering~\cite{Jdong2024}. Sb-based layers have also been identified as promising \textit{n}-type thermoelectric materials, exhibiting an increasing power factor with temperature~\cite{Bafekry2021}.

In the search for potential TE materials, high $ZT$ is often found in materials containing heavy elements such as Pb~\cite{Mamani2024, SHENG2020, YGan2021, STANG2022, CDing2023, STang2025} and Bi~\cite{JWU2019, Wning2022, Jwen2024, D4MA00924J, Varjovi2021, GuoXiong2024}. For example, in Pb$_2$Sb$_2$S$_5$, strong anharmonic phonon interactions lead to shorter phonon lifetimes and lower thermal conductivity, improving $ZT$~\cite{CDing2023}. Similarly, in PbX$_2$ (X = S, Se, Te) monolayers, enhanced phonon scattering due to increased phonon anharmonicity reduces $\kappa_l$, boosting $ZT$ up to 2~\cite{PJin2022}. Two-dimensional monolayers composed of chalcogen combinations such as PbTe, SnTe, and GeTe exhibit inherently low $\kappa_l$ due to strong interactions between acoustic and low-lying optical phonon modes~\cite{Snair2023}. Furthermore, opto-thermal Raman spectroscopy has experimentally verified that the in-plane $\kappa_l$ of GeSe is 2.3~\klunit{} at room temperature~\cite{JPark2024}.

As an alternative to chalcogen-based monolayers, novel 2D layers such as BiYZ (Y $\neq$ Z = Te, Se and S)~\cite{D4MA00924J}, ATeI (A = Sb and Bi)~\cite{SDGuo2017}, BiTeCl~\cite{CHAUHAN2022}, AsSBr~\cite{MLIU2022}, AsTeX (X = Cl, Br and I)~\cite{Poonam2023}, SbYZ (Y = S and Se and Z = Cl, Br, and I)~\cite{D4TA02974G}, and group III-VI monolayers of the form XY (X=B, Al, Ga, In, Tl; Y= O, S, Se, Te, Po)~\cite{Haldun2021} with enhanced $ZT$ have been theoretically proposed for TE applications. Motivated by these observations, the modeling and development of Pb-halogen-based 2D layers for TE applications warrant further theoretical investigation. Rather than focusing on the symmetrical structure of PbX$_2$ (X = S, Se, Te)~\cite{PJin2022}, we explore Pb-based halide JLs with distinct atomic coordination. In this context, the chemical formula of PbXY is chosen, where X and Y are different halogens that satisfy the condition X$\neq$Y. The fundamental energetics and stability of PbXY JLs have been investigated and presented in Ref.~\cite{AES2024}, revealing these JLs as promising candidates for water-splitting applications. Furthermore, the remarkably low thermal conductivity of these JLs, (Sec.~\ref{sec:tc}), suggests their potential as thermoelectric materials.

Having presented the basic properties of PbXY JLs, this study further explores their applications as energy harvesting materials. A comprehensive overview of the thermoelectric properties of Pb-halogen JLs using first-principles calculations combined with the Boltzmann transport equation is presented. The following sections outline the computational details and elaborate on the electronic properties, thermal conductivity, and Onsager coefficients. Based on the deformation potential theory, the carrier scattering rate is calculated and integrated with the Onsager coefficients to estimate $ZT$. The results highlight the potential of PbXY JLs as promising candidates for thermoelectric applications.

\section{\label{method}Methodology}
Spin-polarized \textit{ab initio} simulations based on density functional theory (DFT) are performed using the VASP code~\cite{kresse1996, Kresse1996_prb}. The exchange-correlation interactions are treated within the Perdew-Burke-Ernzerhof (PBE) functional under the generalized gradient approximation (GGA)~\cite{pbe}. Electron-ion interactions are modeled using the projector-augmented wave (PAW) method~\cite{Blochl1994, Kresse1999}, with a plane-wave cutoff energy of \SI{500}{\electronvolt}. The convergence criteria are set to \SI{e-6}{\electronvolt\per\angstrom} for atomic forces and \SI{e-3}{\electronvolt} for total energy. The lattice vectors and atomic arrangements of PbXY JLs follow those reported in Ref.~\cite{AES2024}, see Fig.~\ref{fgr:ballstick}(a)-(b) for the generalized visualization of PbXY JLs. Structural relaxation and ground-state energy calculations are carried out using the Monkhorst-Pack scheme~\cite{monkhorst1976} with a $16\times16\times1$ \textbf{k}-point grid. A vacuum spacing of \SI{15}{\angstrom} is applied along the perpendicular direction to the layer to prevent interactions between periodic images.

To assess the thermal stability of PbXY JLs under ambient conditions, ab initio molecular dynamics (AIMD) simulations are conducted at a fixed temperature of 300~K. The simulations use a supercell with lattice vectors extending $4a_0$ along both the $\vec{a}$ and $\vec{b}$ directions, where $a_0$ is the lattice parameter of the JL unit cell, see Fig.\ref{fgr:ballstick}(c). The system is initially heated for \SI{1}{\pico\second}, followed by equilibration for \SI{3}{\pico\second}, with a time step of \SI{0.5}{\femto\second}. The NVT canonical ensemble is employed, with a Nos\'e-Hoover thermostat acting as a heat bath to maintain the target temperature. Following AIMD, the force information from each time step is processed using the Alamode code~\cite{TT2018_jpsj, TT2014_jpcm} to evaluate the phonon contribution to thermal conductivity. The harmonic force constants are extracted for all possible atomic pairs, while cubic anharmonic force constants are computed for interactions up to a distance of 3$a_0$. The thermoelectric properties of PbXY JLs are investigated by computing the Onsager coefficients using the Boltzmann transport equation (BTE) as implemented in the BoltzTraP code~\cite{btp2}. The band energies from DFT are interpolated using quasi-particle energies and their derivatives, see Eq.(1) in Ref.~\cite{btp2}. Assuming a rigid-band approximation, \ie the band structure remains unchanged with temperature or doping, the quasi-particle energies are analyzed to solve the BTE under the constant relaxation time approximation (CRTA). Within this framework, the electrical conductivity $\sigma$, Seebeck coefficient $S$, and charge-carrier thermal conductivity $\kappa_e$ are determined as

\begin{numparts}\label{eq:bte}
    \centering
    \begin{eqnarray}
        \sigma   ~&=~ \mathcal{L}^{(0)}, \label{eq:sigma}\\
        S        ~&=~ \frac{1}{q\cdot T} \left[ \frac{\mathcal{L}^{(1)}}{\mathcal{L}^{(0)}} \right], \label{eq:S}\\
        \kappa_e ~&=~ \frac{1}{q^2\cdot T} \left[ \frac{(\mathcal{L}^{(1)})^2}{\mathcal{L}^{(0)}} - \mathcal{L}^{(2)} \right], \label{eq:kappa}
    \end{eqnarray}
\end{numparts}
where $q$ and $T$ are the carrier charge and temperature, respectively, and $\mathcal{L}^{(\alpha)}$ represents the generalized transport coefficients as a function of chemical potential $\mu$ and temperature as

\begin{equation}
    \label{eq:L}
        \mathcal{L}^{(\alpha)}(\mu,T) ~=~ q^2 \int \Sigma(\varepsilon, T) (\varepsilon - \mu)^\alpha \cdot\left( \frac{ - \partial f^{(0)}(\varepsilon, \mu, T)}{\partial \varepsilon} \right)  d\varepsilon ,
\end{equation}
where the distribution function $\sigma(\varepsilon, T)$ is then obtained using the linearized BTE under the CRTA as,

\begin{equation}
    \label{eq:df}
        \Sigma(\varepsilon, T) ~=~ \int \sum_b \mathbf{v}_{b,\bm{k}} ~\otimes ~\mathbf{v}_{b,\bm{k}} ~\tau_{b,\bm{k}} ~\delta(\varepsilon-\varepsilon_{b, \bm{k}}) \frac{d\bm{k}}{8\pi^3}
\end{equation}
More details about $\mathcal{L}^{(\alpha)}(\mu,T)$ and its implementations can be obtained from Ref.~\cite{btp2, btp}.

\section{Results}
\subsection{Structural properties}
\begin{figure*}[tb]
 \centering
 \includegraphics[width= 1\textwidth]{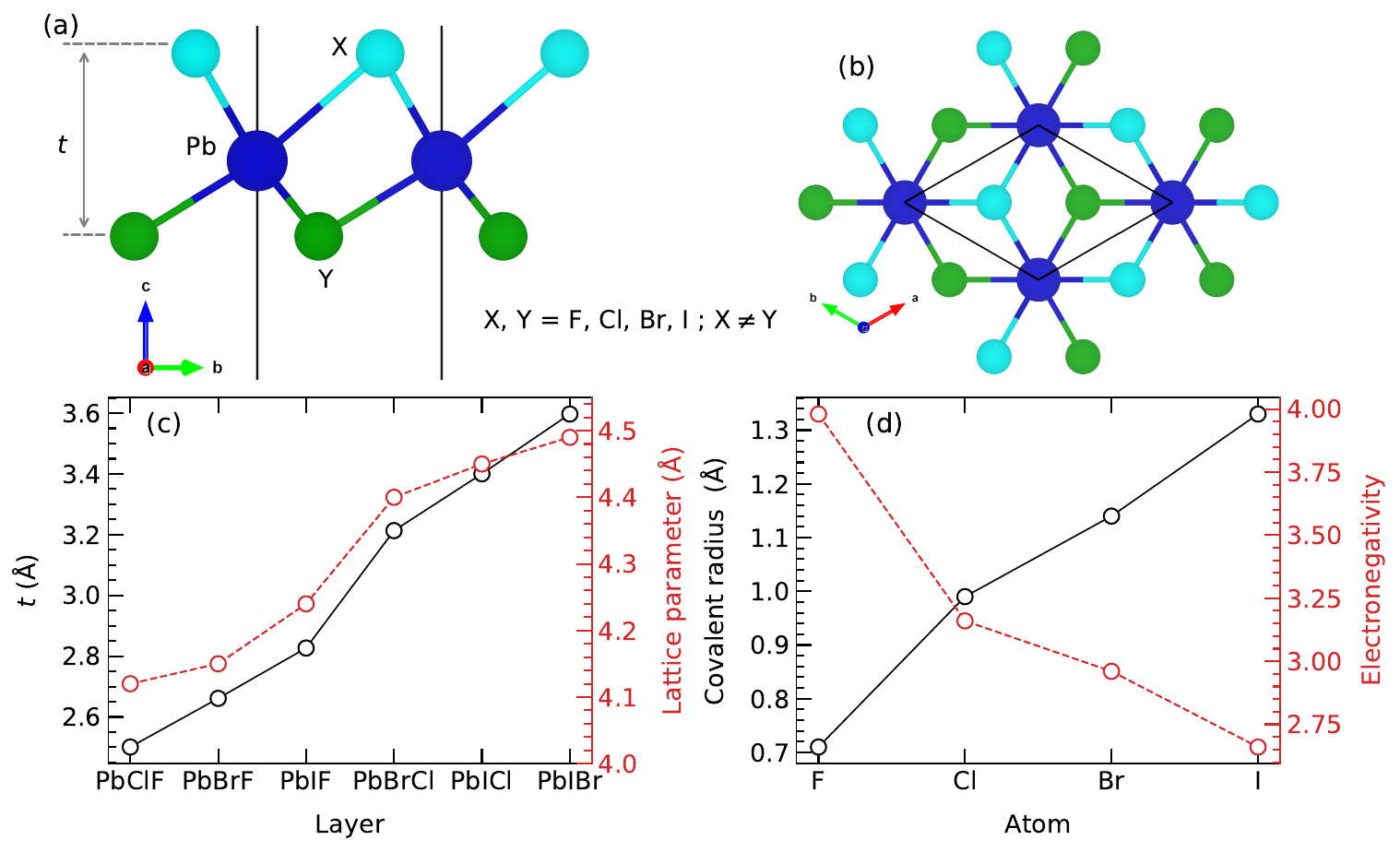}
 \caption{The general ball-and-stick representation of PbXY JLs is shown in a) cross-sectional and b) perpendicular views, cf. Fig.~1 in Ref.~\cite{AES2024} for individual layer arrangement. The thickness $t$ is calculated between the $ X$-$Y$ sublayers for each JL and is shown in (c) along with the optimized lattice parameter. For comparison purposes, the electronegativity and covalent radius of constituent halides are given in (d). The increasing atomic size of halides increases the lattice parameter, and the thickness $t$ increases due to the combined effect of electronegativity and atomic size. The atomic views in (a) and (b) are generated using the VESTA package~\cite{vesta}.}
 \label{fgr:ballstick}
\end{figure*}

The general atomic arrangements of PbXY JLs are shown in Fig.~\ref{fgr:ballstick}(a)-(b). The central cationic Pb sublayer, in Pb$^{2+}$ oxidation state, is surrounded by the anionic sublayers X and Y, in X$^{1-}$ and Y$^{1-}$ oxidation states. The X-Pb-Y sandwich atomic arrangements of JLs lead to the $C_{3v}$ point group with the non-centrosymmetric $P3m1$ space group, regardless of X and Y combinations. Though the oxidation states of halides are the same, their electronegativity $EN$ determines the nature of Pb-X and Pb-Y bonding and hence their bond lengths, thereby inducing different electronic characteristics. In general, $EN$ of Y is always kept higher than that of X. Therefore, the Pb-Y bond length is always shorter than that of Pb-X, cf. Tab.~1 in Ref.~\cite{AES2024}. The difference in $EN$ $\Delta EN$ can indicate a rough estimation of the percentage of ionic character, known as ionicity of the bonding by $Ionicity~\text{(in \%)} = 100\times\left(1-e^{-(0.5\times\Delta EN)^2}\right)$~\cite{Xiaoli2025}. The calculated ionicity of Pb-F, Pb-Cl, Pb-Br, and Pb-I is approximately 49.4\%, 15.8\%, 9.4\%, and 2.7\%, respectively. These observations elucidate that the ionicity decreases as the size ($EN$) of halides increases (decreases), \ie, in general, the bonds in JLs containing F atoms are more ionic than those in other JLs, as the $EN$ of F is larger than that of any other element. 

Also, the thickness $t$ of JLs, defined as the distance between the halide sublayers, is correlated to the lattice parameter of the layers, cf. Fig.~\ref{fgr:ballstick}(c). Figure~\ref{fgr:ballstick}(c) and (d) demonstrate that the increasing trend in $t$ is a direct consequence of both the atomic size and electronegativity of halides. As the atomic size of halides (the covalent radius) increases, the layer expands in all crystallographic directions and $t$ increases. In addition, the changes in $t$ are reflected in the lattice parameter. The JL with the larger halides (PbIBr) has a larger unitcell vector than other JLs. The observed trend of $t$ elegantly classifies the PbXY JLs into two categories: the JLs (i) consisting of the smallest halide, the F atom, \ie PbClF, PbBrF, and PbIF (hereafter mentioned as $F$-JL) and (ii) consisting of a mixture of larger halides \ie PbBrCl, PbICl, and PbIBr (hereafter mentioned as $ L$-JL). Interestingly, in $F$-JLs, the Pb-F bond length remains the same ($2.5$~\AA), whereas the Pb-Y bond length increases as halide size increases. In general, the bond lengths in PbXY JLs can be correlated as $l_\mathrm{Pb-F}~<~l_\mathrm{Pb-Cl}~<~l_\mathrm{Pb-Br}~<~l_\mathrm{Pb-I}$, irrespective of the category.  

\subsection{\label{sec:elecpro}Electronic properties}
The effect of the EN difference between the halides in JLs straightforwardly alters the characteristics of holes and electrons in $F$-JLs. For comparison, the energy levels around the Fermi energy are presented for each JL in SFig.~1 in the SM. When mixed with other halides, F attracts more electrons and forms strong bonds with Pb, equivalently, leading to the smallest $l_\mathrm{Pb-F}$. Such interaction accumulates more charge between Pb and F, accompanied by a stronger field, which may increase the effective mass of charge carriers, both electrons and holes. Therefore, the energy levels in the valence region around the Fermi level have distinct curvature, varying depending on the X and Y electronegativity ratio. The variation in the curvature shifts the valence band maximum away from the high symmetry location $K$ and broadens the conduction band while shrinking the band gap. The Br-F combination intriguingly introduces a more subtle maximum at the valence band, cf. SFig. 1(b). The Br and F electronegativity ratio acts as a critical limit, beyond which the curvature of the valence band can be either positive or negative, cf. SFig. 1(a)-(c). These curvature changes result in a peculiar direct bandgap exclusively for the PbIF JL, whose valence curvature switches to negative at $\Gamma$, $\ie$ the energy bands are parabola facing down. Unlike $F$-JLs, $L$-JLs do not expose such curvature change %, and the valence band's profile is almost more intact 
upon changing the halide combination. % All $L$-JLs show negative curvature at the valence band, $\ie$, the parabolic energy band is facing down. 
Therefore, the $L$-JLs possess an indirect bandgap. 

Along with variation in bandgap type, the valence band flattens along the $\Gamma - K$ direction and presents carrier pockets, \ie the abrupt dispersion of energy bands over short reciprocal path lengths, along the $\Gamma - M$ direction, similar to that in PbTe$_{1-x}$Se$_{x}$~\cite{Pei2011} and $\alpha$-Te monolayer~\cite{ZGao2018}. Such a blending is a manifestation of exemplary TE properties~\cite{Pei2011, ZGao2018}. Therefore, the changes in the conduction and valence bands are expected to influence the electronic conductivity and carrier mobility of the intended JLs and control the thermoelectric power factor.

\subsection{Carrier mobility and Relaxation time}
\begin{figure}[tb]
 \centering
 \includegraphics[width=1.0\textwidth]{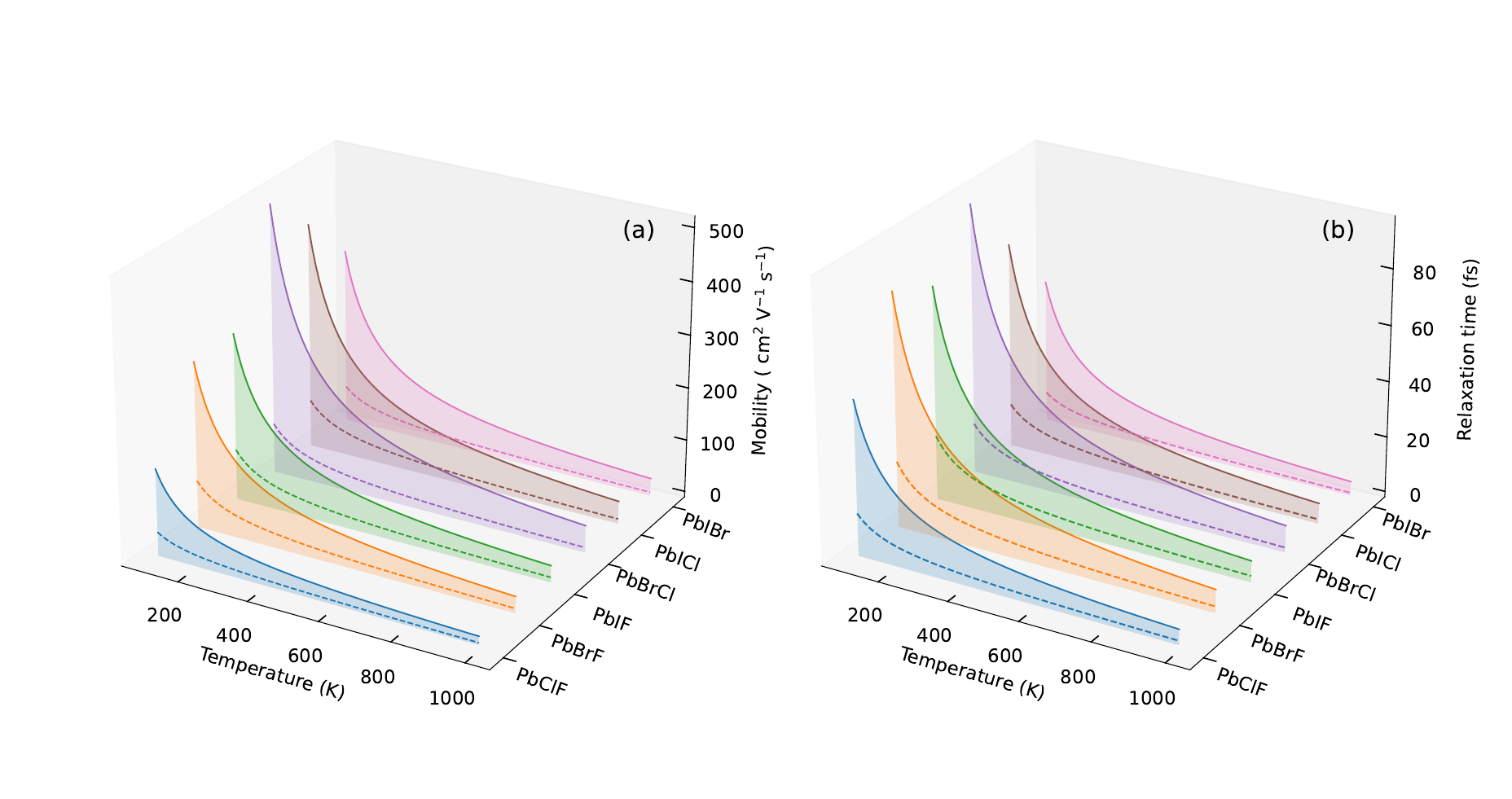}
 \caption{The calculated $x$ and $y$-components of (a, c) mobility $\mu$ and (b, d) corresponding relaxation time $\tau$ of electrons in PbXY JLs. The electrons generally present moderate mobility and require up to 1000 fs to relax. The hole carriers relax faster than electrons. Note that only a negligible carrier mobility difference is observed between PbBrCl and PbIBr JLs.}
 \label{fgr:e_mu-tau_x}
\end{figure}

The relaxation time $\tau$ is calculated by applying the deformation potential theory (DP)~\cite{Bardeen1950, Xi2012}, proposed by Bardeen and Shockley. Under the DP approximation, the $\tau$ is defined as~\cite{Mamani2024, Xi2012, kaur2019} 
\begin{equation}
    \label{eq:tau}
        \tau ~=~  \frac{\hbar^3 \, C_\beta}{k_B \, T \, m^* \, m_d \, E_{\beta}^{2}},
\end{equation}
where the parameters $C_\beta$, $E_{\beta}$, $m^*$, and $m_d$ represent the elastic constant, deformation potential, effective mass of carriers, and average effective mass of carriers, respectively. The elastic constant is calculated directly from the first-principles approximation, and the average effective mass $m_d$ is determined from the effective mass along $a$ and $b$ lattice vectors as $\sqrt{m_x^* m_y^*}$. The deformation $E_{\beta}$ is calculated as the ratio between the change in the energy of conduction (valence) band minimum (maximum) against the variation in cell size, \ie $E_{\beta}~=~\nicefrac{\Delta E_\mathrm{{CBM/VBM}}}{\Delta l/l_0}$. Then the mobility $\mu$ of charge carriers is derived as $\mu \, = \, \tau e / m^*$. The calculated $\mu$ and relaxation time for electrons is shown in Fig.~\ref{fgr:e_mu-tau_x}, for holes see Fig. 5 in SM.

Generally, the carriers in $F$-JLs are less mobile than those in $L$-JLs. As explained above, the carrier pockets are observed only in $L$-JLs valence bands, which improve $\mu$. The flattened energy bands of $F$-JLs, cf. SFig. 1 in SM, severely affect the charge carriers, limiting the mobility. The carrier packets in energy bands of $L$-JLs accelerate the charge carriers, and the mobility ranges from a few hundred to a thousand. As an example, the mobility of charge carries reaches \SI{594}{} cm$^2$ V$^{-1}$ s$^{-1}$ and \SI{634}{} cm$^2$ V$^{-1}$ s$^{-1}$ at \SI{300}{\kelvin} in PbBrCl and PbIBr JLs, respectively and to \SI{134e4} cm$^2$ V$^{-1}$ s$^{-1}$ at 300 K in PbICl JL, higher than the reported values for other 2D materials such as MoS$_2$ (104 cm$^2$ V$^{-1}$ s$^{-1}$ and 165 cm$^2$ V$^{-1}$ s$^{-1}$ for holes and electrons respectively)\cite{VHa2024npj}. Also, the electron has a higher mobility than the holes in the complete range of intended temperatures, cf. hole mobility in SFig. 5 in SM. The flat valence bands and the absence of carrier packets in the valence band strongly affect the mobility of holes. Further improvement in mobility can be achieved by generating more carrier packets, substituting suitable dopants, or using an appropriate substrate material. For example, the transistors fabricated using MoS$_2$ thin films exhibit nearly a twofold enhancement in carrier mobility from 9.1 cm$^2$ V$^{-1}$ s$^{-1}$ to 19.9 cm$^2$ V$^{-1}$ s$^{-1}$, when supported by a 20 nm thick HfO$_2$ substrate~\cite{Huang2023APL}. 

The calculated relaxation time $\tau$ of charge carriers reveals key differences between the $F$- and $L$-JLs. In general, electrons and holes exhibit longer $\tau$ along the zigzag direction (denoted as $x$-direction in Fig.~\ref{fgr:e_mu-tau_x}) than along the armchair direction (denoted as $y$-direction in Fig.~\ref{fgr:e_mu-tau_x}). Also, the $\tau$ of electrons is longer than that of holes, cf. $\tau$ at ~300 K in Table S1 and S2 in SM. The longer $\tau$ hints at a lesser scattering, and the carrier can move faster. Therefore, enhanced electrical conductivity is expected in PbICl due to high mobility and relaxation time.

\subsection{\label{sec:tc}Phonon thermal conductivity}
\begin{figure}[tb]
 \centering
 \includegraphics[width= 1\textwidth]{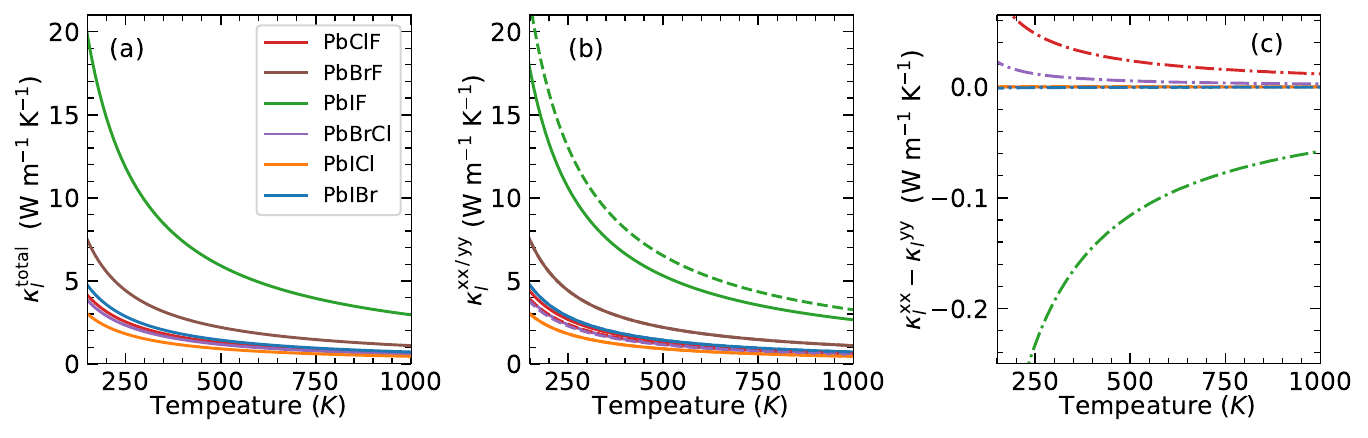}
 \caption{The calculated a) total phonon contributed thermal conductivity $\kappa_l^{total}$, b) component resolved $\kappa_l$ along $xx$ ($\kappa_l^{xx}$, solid lines) and $yy$ ($\kappa_l^{yy}$, dashed lines), and c) components difference of $\kappa_l$ ($i.e.$ $\kappa_l^{xx}-\kappa_l^{yy}$). The $F$-JLs (PbIF and PbClF) have a larger directionally dependent $\kappa_l$, and the difference is almost negligible for other $F$- and $L$-JLs. The remaining components of $\kappa_l$ have vanished in the entire spectrum of the intended temperature.}
 \label{fgr:kl-total}
\end{figure}

Following the relaxation time of charge carriers, the phonon–phonon interaction is analyzed to calculate the lattice thermal conductivity $\kappa_l$, which is crucial in maximizing the $ZT$. Achieving a low $\kappa_l$ as much as possible enhances the TE performance of materials. The $\kappa_l$ is calculated by combining the AIMD with the anharmonic lattice dynamics (ALD) method~\cite{TT2014_jpcm}. In AIMD, all atoms in the simulation cell are displaced from their equilibrium positions correlated with the target temperature. The displacement magnitude $u$ of individual atoms at each pico-second AIMD iteration can be extracted relative to the equilibrium position at 0 K. This displacement pattern is combined with the respective force information to frame the displacement-force dataset from AIMD, providing accurate harmonic and higher-order interatomic force constants. The $\kappa_l$ is calculated using this dataset based on Boltzmann transport theory within the ALD framework as~\cite{TT2014_jpcm, kaur2019, Siddi2024}  

\begin{equation}\label{eq:kappal}
\kappa _{l}^{\mu ,\nu} = \frac{1}{V N} \sum_{\boldsymbol{q}, j} c\left( \boldsymbol{q}, j \right) v^{\mu}\left(\boldsymbol{q}, j\right) v^{\nu}\left(\boldsymbol{q}, j\right) \tau \left(\boldsymbol{q}, j\right),
\end{equation}
where $V$ and $N$ are the volume of the unit cell and the number of $q$ points in the first Brillouin zone, respectively. The summations over the phonon branches $j$ and wave vectors $\boldsymbol{q}$ include the contributions of constant-volume specific heat $c\left(\boldsymbol{q}, j\right)$, the group velocity $v\left(\boldsymbol{q}, j\right)$, and the relaxation time $\tau \left(\boldsymbol{q}, j\right)$ of phonons. The indices $\mu$ and $\nu$ represent Cartesian components of the indexed quantities. As common practice for 2D materials~\cite{Ould2024}, the computed $\kappa_l$ is multiplied by $z/t$, where $z$ is the length of the unitcell along crystallographic direction $c$ and $t$ is the thickness of the JLs. The calculated total $\kappa_l$ (\ie average of components), the component resolved $\kappa_l$, and the anisotropy in $\kappa_l$ components $\Delta \kappa_l ~=~\kappa_l^{xx}-\kappa_l^{yy}$ are shown in Fig.~\ref{fgr:kl-total}. 

% PbClF    2.11961
% PbBrF    3.674808
% PbIF     9.855757
% PbBrCl   1.942471
% PbICl    1.511433
% PbIBr    2.38306
Among all the JLs, PbIF exhibits the highest $\kappa_l$ and the largest variation in $\kappa_l$ across the studied temperature range. PbBrF shows a moderate $\kappa_l$, while the remaining JLs have comparable values of approximately $\approx3$\klunit{} at 300 K. In all cases, $\kappa_l$ decreases with increasing temperature, consistent with enhanced phonon-phonon scattering. Notably, this temperature dependence is more pronounced in the low-temperature regime, and above 250 K $\kappa_l$ scales as $T^{-1}$, similar to theoretical observations in Pb-containing bulk samples such as Pb$_2$Sb$_2$S$_5$, Pb$_2$Sb$_2$Se$_5$, and Pb$_2$Sb$_2$Te$_5$~\cite{YGan2021}. The differences in $\kappa_l$ among the studied JLs can be rationalized by analyzing their phonon band structures, see Fig. S6, where clear distinctions emerge. In general, low (high) frequency optical phonons are contributed by the vibration of lighter (heavier) halide constituents, and the acoustic phonon frequencies originate from the heavier cation Pb$^{2+}$. The low-frequency optical modes in $F$-JLs overlap with acoustic modes, particularly the first two. This overlapping becomes progressively stronger from PbClF to PbBrF to PbIF, attributed to the atomic mass and size of $X$ and $Y$ atoms. The mass difference between $X$ and $Y$ accounts for interaction between optical phonons, $\ie$ an emerging gap between optical branches, as observed in bulk crystals of $F$-JLs~\cite{yedukondalu2022}. In contrast, the degree of optical–acoustic mode crossing diminishes in $L$-JLs as the atomic mass and size disparity between atoms $X$ and $Y$ decreases. Moreover, mode switching between acoustic and optical branches is evident in the $F$-JLs and most prominent in PbIF, resulting in strong acoustic–optical hybridization and the participation of more optical modes in thermal transport. This accounts for the significantly higher $\kappa_l$ of PbIF than that of other JLs. Additionally, a distinct dispersive hump appears in the second optical branch near the Brillouin zone center, with flat curvature in $F$-JLs and more dispersive curvature in $L$-JLs along the $\Gamma$–$M$–$K$–$\Gamma$ direction. This dispersion, coupled with mode repulsion, leads to a suppression of $\kappa_l$ in $L$-JLs.

\subsection{\label{sec:ph-lt}Phonon life time}
\begin{figure}[tb]
 \centering
 \includegraphics[width= 1\textwidth]{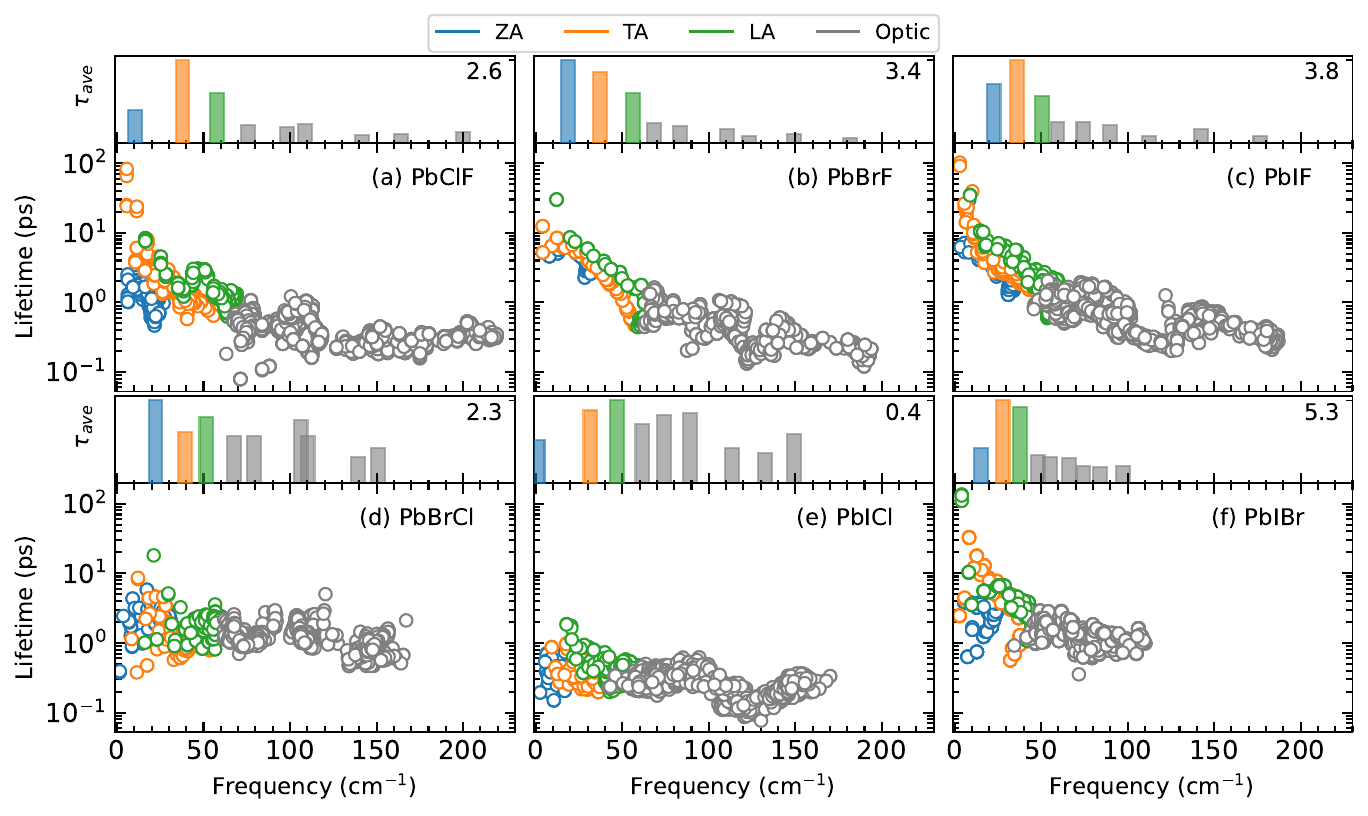}
 \caption{The mode-resolved phonon lifetime $\tau$ as a function of frequency in PbXY JLs, derived using the ALD method. The bars in the upper boxes show the average lifetime of individual phonons as a function of their average frequency. The number in a box gives the maximum average lifetime (highest bar). Comparison of bars concludes that the optical phonon modes have a shorter life span than the acoustic modes in $F$-JLs. However, the life span of optical phonons is comparable to that of the acoustic phonons in PbBrCl and PbICl JLs.}
 \label{fgr:phonlt}
\end{figure}
The curvature of phonon bands provides insight into the heat-carrying dynamics of the layers. Each phonon mode contributes individually to $\kappa_l$, with longer-lived phonons making more substantial contributions. This contribution can be assessed through the phonon lifetime (or relaxation time) $\tau\left(\boldsymbol{q}, j\right)$, which is presented in Fig.~\ref{fgr:phonlt} with the average phonon lifetime $\tau_{\mathrm{ave}}$ as a function of the average frequency of individual modes. The $\tau_{\mathrm{ave}}$ represents the temporal persistence of the modes, while the mean frequency (indicated by the bar positions) qualitatively reflects the contribution of modes to $\kappa_l$. As shown in Fig.~\ref{fgr:phonlt}, optical modes in the $F$-JLs exhibit significantly shorter lifetimes than acoustic modes, indicating their weaker role in heat conduction~\cite{MinLi2024prb}. In particular, transverse acoustic (TA) phonons dominate the thermal transport in PbClF and PbIF, with acoustic frequencies extending up to 60 cm$^{-1}$. For PbBrF, out-of-plane acoustic (ZA) phonons primarily govern heat transport with acoustic frequencies spanning up to 55 cm$^{-1}$. In contrast, the optical phonons exhibit relaxation times comparable to acoustic modes in $L$-JLs. Such optical phonons have a stronger influence on thermal transport. This results in a combined acoustic–optical contribution to $\kappa_l$ in the $L$-JLs.

Among the considered JLs, PbICl and PbIBr exhibit the minimum and maximum average phonon lifetimes $\tau_{\mathrm{ave}}$, respectively. Layers PbBrCl, PbIBr, and PbClF show nearly similar $\kappa_l$, which can be attributed to a balancing act between phonon group velocity and lifetime. Although PbICl features the acoustic phonons with frequency range extending up to 55 cm$^{-1}$, comparable to that of the $F$-based JLs, the significantly lower $\tau_{\mathrm{ave}}$ (about an order of magnitude smaller than in other JLs) indicates enhanced phonon scattering, which suppresses the $\kappa_l$ significantly. The PbClF and PbBrCl JLs have comparable thermal conductivities due to their similar $\tau_{\mathrm{ave}}$ and phonon frequency ranges. While having the highest $\tau_{\mathrm{ave}}$ among the JLS, the lower frequency span in PbIBr JL limits the effective heat transport, leading to a $\kappa_l$ comparable to those of PbClF and PbBrCl JLs.

\subsection{\label{sec:grueneisen}Gr\"uneisen parameter}
\begin{figure}[tb]
 \centering
 \includegraphics[width= 1\textwidth]{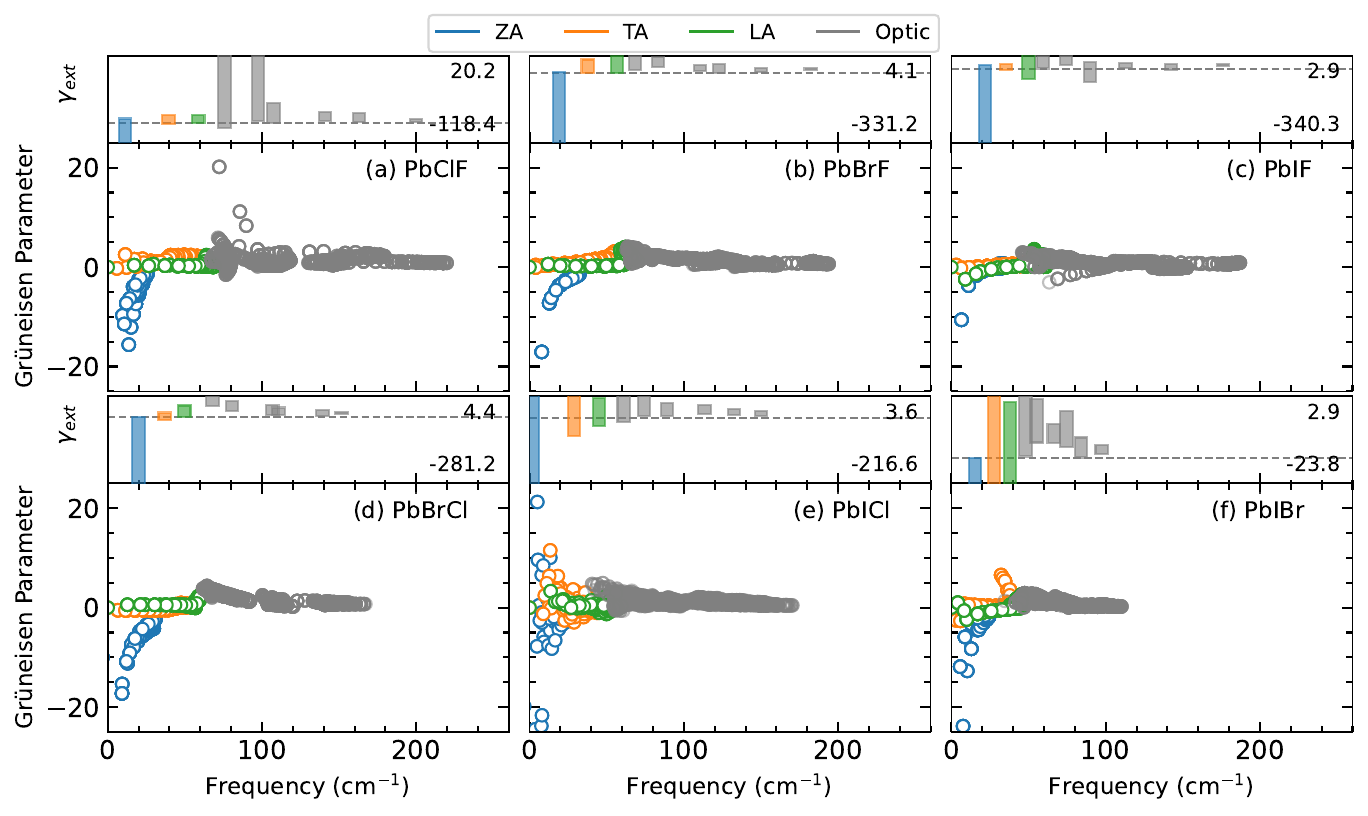}
 \caption{Mode resolved Gr\"uneisen parameter $\gamma$ as a function of frequency in PbXY JLs. The bars in the upper boxes represent the range of $\gamma$ values for individual modes as functions of the average frequency of the modes. The values in the boxes give the minimum and maximum of $\gamma$, and the dashed line represents $\gamma=0$. Overall, the ZA mode has a negative correlation in all JLs except PbICl, in which both positive and negative correlation is observed.}
 \label{fgr:grueneisen}
\end{figure}

The aforementioned acoustic-optical phonon interactions induce strong phonon scattering, significantly reducing the lattice thermal conductivity~\cite{Mamani2024}. The impact of lattice vibrations on phonon scattering can be quantitatively characterized by the dimensionless Grüneisen parameter $\gamma$~\cite{Xoiong2025, D4MA00924J, D4TA02974G, Nicola2005}, which is defined as~\cite{TT2014_jpcm}
\begin{equation}\label{eq:grueneisen}
\gamma_{j}(\boldsymbol{q}) = \frac{V}{\omega_{j}(\boldsymbol{q})} \frac{\partial \omega_{j}(\boldsymbol{q})}{\partial V},
\end{equation}
where $j$ is the phonon mode index, $V$ the simulation cell volume, and $\omega_{j}(\boldsymbol{q})$ denotes the phonon frequency of mode $j$ at wavevector $\boldsymbol{q}$. The mode-resolved Gr\"uneisen parameters are shown in Fig.~\ref{fgr:grueneisen}, along with the maximum and minimum values as a function of average frequency of each phonon mode. Large $\gamma$ values imply strong anharmonicity, which enhances phonon-phonon scattering and reduces phonon lifetimes, thereby suppressing the lattice thermal conductivity~\cite{YGan2021, Dfan2017}. The sign of $\gamma$ indicates the response of phonon frequency to lattice volume changes, as the positive $\gamma$ denotes frequency softening upon expansion, while a negative $\gamma$ signifies frequency hardening. This distinction reflects the asymmetric phonon behavior under tensile and compressive strain, revealing mode-specific anharmonic interactions~\cite{peng2019jap, Ye2024prb}.

Moreover, the mass contrast between the halide constituents $X$ and $Y$ influences the variation of $\gamma$ in $F$-JLs, where the range of the $\gamma$ values increases systematically from PbClF to PbBrF to PbIF, following the increasing mass difference. A similar trend is observed in $L$-JLs, as seen in PbICl and PbIBr. This mass-difference-driven effect on $\gamma$ has also been reported in both 2D materials (such as Pb$_2$Sb$_2$S$_5$, Pb$_2$Sb$_2$Se$_5$, and Pb$_2$Sb$_2$Te$_5$)~\cite{YGan2021} and bulk crystals (such as BiCuSeO and LaCuSeO)~\cite{Dfan2017}.

The mode-resolved $\gamma$ analysis depicts that the ZA modes exhibit a negative correlation with volume in all JLs, except for PbICl, which shows both positive and negative values. These negative $\gamma$ indicate phonon hardening under lattice expansion, which is a characteristic behavior of layered materials arising from their flexural nature and weak interlayer interactions, enhancing anharmonicity in ZA modes~\cite{peng2019jap, Ye2024prb, yangpccp2021}. Interestingly, the ZA mode in PbICl shows a mixed character, suggesting a more complex vibrational response. This complexity is likely influenced by the significant electronegativity and charge difference between I and Cl atoms in PbICl, which alters the local bonding environment more than that in other JLs~\cite{AES2024}. 

The in-plane acoustic modes TA and LA consistently show positive $\gamma$ in all JLs, indicating typical phonon softening upon lattice expansion~\cite{Ye2024prb}. Similarly, the optical modes exhibit positive $\gamma$, though with smaller magnitudes than acoustic modes. This suggests a weak contribution of optical modes to $\kappa_l$, consistent with their short lifetime $\tau$. These trends underscore the dominant role of acoustic phonons in determining the $\kappa_l$ of the layers. A similar conclusion can be derived by other parameters such as phonon scattering rate, group velocity, and mean free path, which are shown in SFig. 2, 3, and 4, respectively, in SM. The intrinsically low $\kappa_l$ supports the potential for achieving high thermoelectric performance in these layers. 

Overall, the strong phonon scattering rates and large Gr\"uneisen parameters in both $F$- and $L$-JLs highlight their pronounced anharmonicity and potential for thermoelectric applications. Among them, PbICl stands out due to its distinct scattering characteristics. The intrinsically low $\kappa_l$ in these JLs further reinforces their promise for high-performance thermoelectric devices.

\subsection{Thermoelectric properties}
\begin{figure}[tb]
 \centering
 \includegraphics[width= 1\textwidth, clip, trim=0.1in 0.3in 0.1in 0.0in]{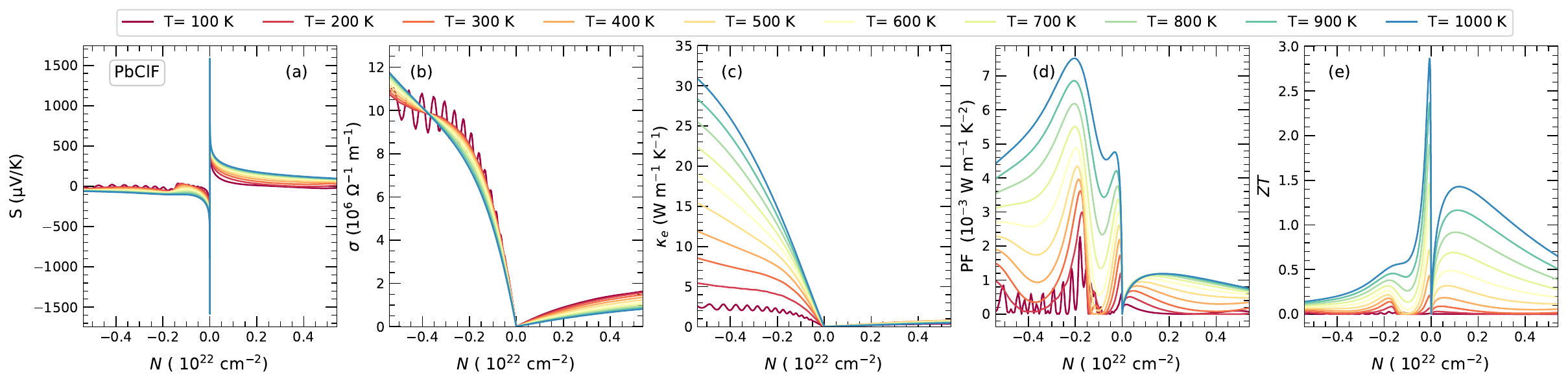}
 \includegraphics[width= 1\textwidth, clip, trim=0.1in 0.3in 0.1in 0.4in]{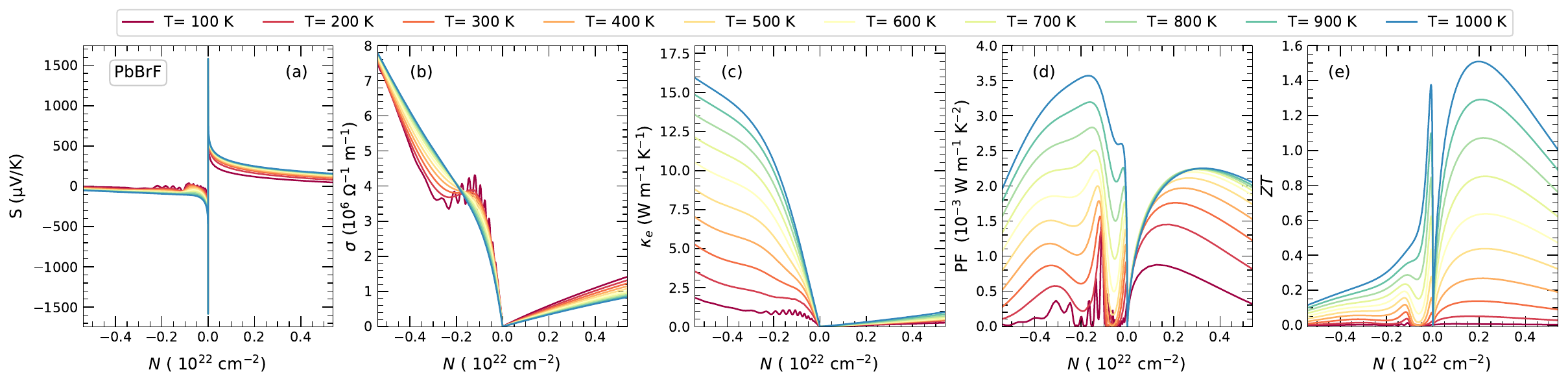}
 \includegraphics[width= 1\textwidth, clip, trim=0.1in 0.0in 0.1in 0.4in]{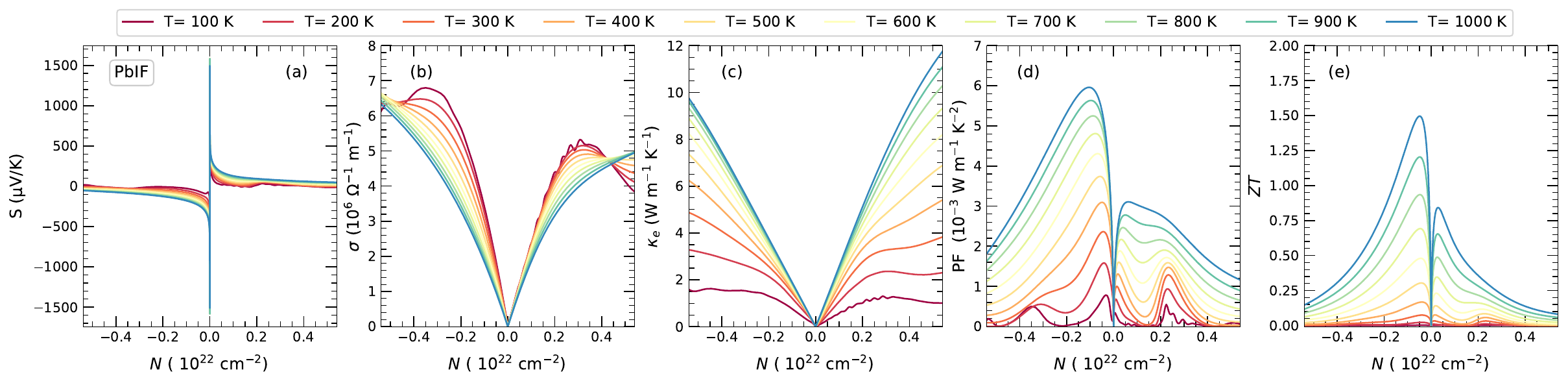}
 \caption{The calculated a) $S$, b) $\sigma$, c) $\kappa_e$, d) $PF$, and e) $ZT$ of PbClF (top row), PbBrF (middle row), and PbIF (bottom row) JLs. The negative (positive) $N$ represents the electron (hole) carrier concentration. The $PF$ of JLs is in the order of $\SI{e-3}{}$ \pfunit{}, higher than many other 2D materials.}
 \label{fgr:TEx-FJL}
\end{figure}

\begin{figure}[tb]
 \centering
 \includegraphics[width= 1\textwidth, clip, trim=0.1in 0.3in 0.1in 0.0in]{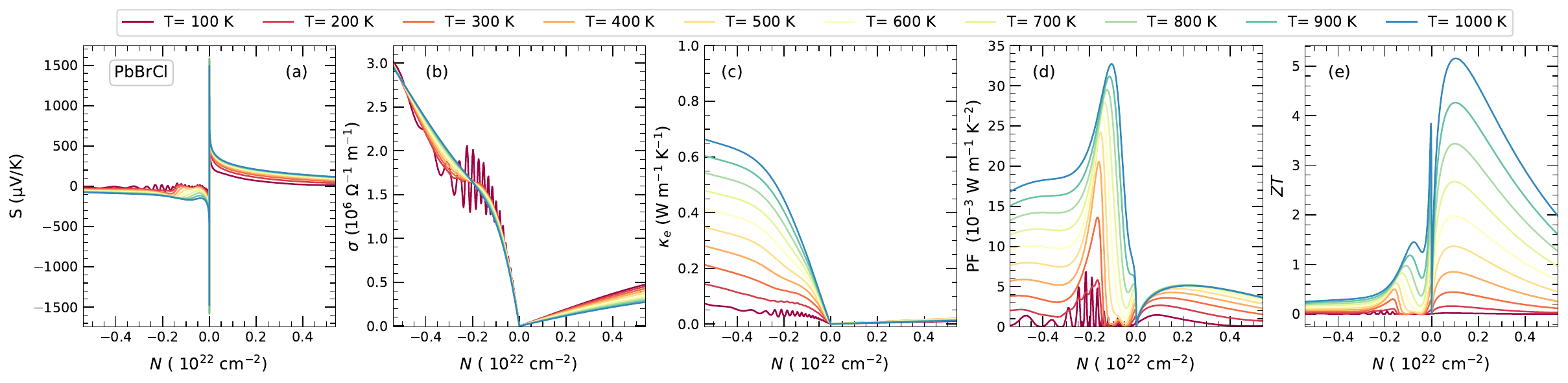}
 \includegraphics[width= 1\textwidth, clip, trim=0.1in 0.3in 0.1in 0.4in]{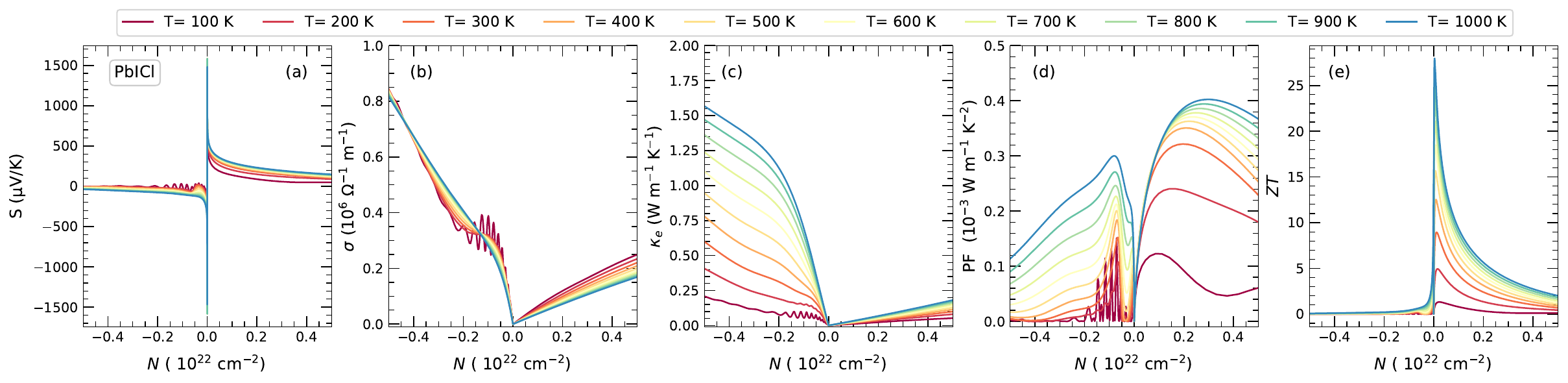}
 \includegraphics[width= 1\textwidth, clip, trim=0.1in 0.0in 0.1in 0.4in]{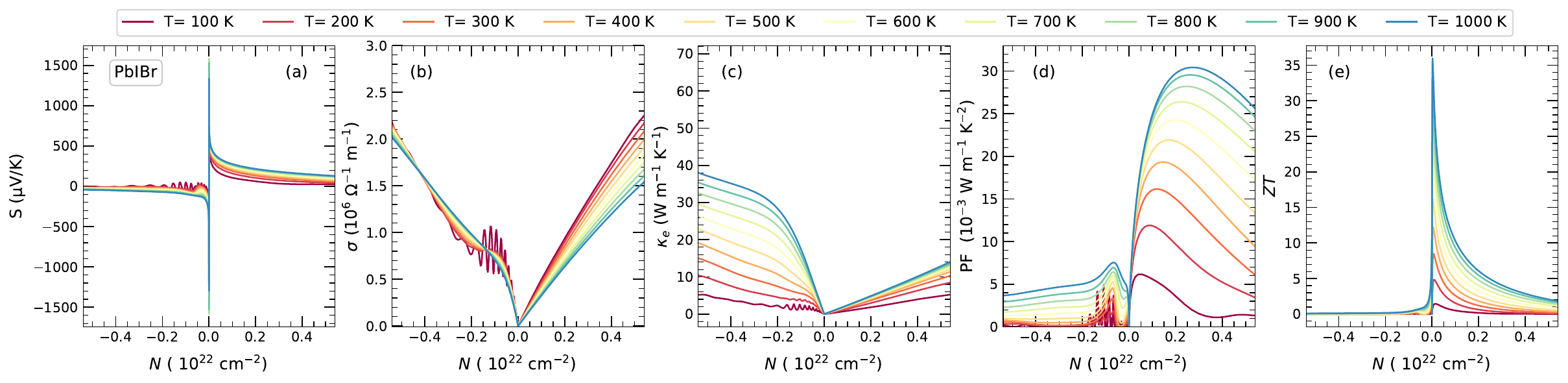}
 \caption{The calculated a) $S$, b) $\sigma$, c) $\kappa_e$, d) $PF$, and e) $ZT$ of PbBrCl (top row), PbICl (middle row), and PbIBr (bottom row) JLs. The negative (positive) $N$ represents the electron (hole) carrier concentration. The $\sigma$ of PbICl is two orders of magnitude higher than that of the JLs, which enhances the $PF$ while that of PbBrCl and PbIBr is in  $\SI{e-3}{}$ \pfunit{}.}
 \label{fgr:TEx-LJL}
\end{figure}

Utilizing the estimated lattice thermal conductivity $\kappa_l$ and the carrier relaxation time $\tau$, the TE properties are computed using the linearized Boltzmann transport equation under the constant relaxation time approximation. In this framework, the electrical conductivity (Eq.~\ref{eq:sigma}) and electronic thermal conductivity (Eq.~\ref{eq:kappa})  obtained from BoltzTrap are normalized by $\tau$, yielding $\sigma/\tau_e$ and $\kappa_e/\tau_e$, respectively. To derive the required $\sigma$ and $\kappa_e$, the relaxation time $\tau$ is calculated based on deformation potential theory, as a standard approach for 2D materials in the absence of experimental data for $\tau$~\cite{Mamani2024, YGan2021, Shafique2017, MILI2022cms}. The parameter elastic constants, deformation potentials, and effective masses used to calculate $\tau$ at 300 K are provided in Tables S1 and S2 of SM. The calculated $\tau$ values exhibit directional dependence, $\ie$, electrons relax at different times along the zigzag and armchair directions. This anisotropy suggests potential differences in TE performance along different directions. Accordingly, the figure of merit $ZT$ is computed separately for each direction using the corresponding $\tau$ values. The TE parameters Seebeck coefficient $S$, electrical conductivity $\sigma$, electronic thermal conductivity $\kappa_e$, and derived power factor $PF~(=S^2\sigma)$, are presented as a function of carrier concentration $N$ in Figs.~\ref{fgr:TEx-FJL} and \ref{fgr:TEx-LJL} for the $F$- and $L$-JLs, respectively. Note that the zig-zag relaxation times of electrons and holes are applied in Figs.~\ref{fgr:TEx-FJL} and \ref{fgr:TEx-LJL}. The armchair relaxation times of electrons and holes are also used to get the TE parameters and are presented in SFig. 7 and 8 for $F$- and $L$-JLs, respectively. The carrier concentration is calculated by setting the Fermi level at the midpoint between the conduction band minimum and valence band maximum, following the convention implemented in BoltzTrap. Therefore, the negative (positive) values of $N$ correspond to electron (hole) doping, obtained by shifting the Fermi level toward the conduction (valence) bands.

The PbXY JLs are semiconductors with sizable bandgap, making them promising candidates for TE applications. The TE efficiency can be enhanced through tuning charge carrier concentration to optimize the TE performance. While heavier charge carriers can increase the Seebeck coefficient, as $S\propto m^*$, they reduce electrical conductivity due to their low mobility~\cite{snyder08_natmat}. Hence, identifying an optimal carrier concentration that balances these contrasting effects is crucial for maximizing the figure of merit $ZT$ and assessing the TE potential of these materials. For practical applications, it is particularly advantageous if the optimal carrier concentration is closer to the Fermi level~\cite{PJin2022}.

Although the Seebeck coefficient is proportional to the bandgap of the material~\cite{gibbs2015}, the maximum absolute value of $|S|$ remains modest due to the minor bandgap variations across the PbXY JLs (cf. SFig. 1 in SM). The magnitude of $|S|$ increases with temperature, consistent with the Mott relation between $|S|$ and temperature~\cite{Mamani2024, Mohebpour2021}. The peak values of $|S|$ are observed within a narrow carrier concentration range of approximately $\SI{\pm0.05e22}{\per\centi\metre\squared}$ around the Fermi level. In both $F$- and $L$-type JLs, the maximum $|S|$ reaches approximately $\SI{1500}{\micro\volt\per\kelvin}$ at $\SI{300}{\kelvin}$, cf. column (a) in Fig.~\ref{fgr:TEx-FJL} and~\ref{fgr:TEx-LJL}. This value surpasses those reported for several other 2D materials, such as Pb$_2$SSe ($\sim\SI{700}{\micro\volt\per\kelvin}$)~\cite{Mamani2024}, PbSnS$_2$ ($\sim\SI{1400}{\micro\volt\per\kelvin}$)~\cite{CDing2023}, and BiSeS ($\sim\SI{1000}{\micro\volt\per\kelvin}$)~\cite{D4MA00924J}. However, it remains lower than that of MoS$_2$ ($\sim\SI{3000}{\micro\volt\per\kelvin}$)~\cite{MV2020}, Si$_2$PH$_2$ ($\sim\SI{2700}{\micro\volt\per\kelvin}$), and Ge$_2$PH$_2$ ($\sim\SI{2800}{\micro\volt\per\kelvin}$)~\cite{Mohebpour2021}.

In addition to $S$, the electrical conductivity $\sigma$ is a key factor in determining the TE efficiency of the layers. A significant variation in $\sigma$ is observed among the $F$-JLs as $N$ and temperature $T$ vary. In PbClF and PbBrF JLs, electron doping enhances $\sigma$, whereas the effect of hole doping is much smaller, $\ie$ $\sigma$ does not increase so rapidly with $N$. This contrasting behavior can be attributed to the difference in effective masses of electrons and holes (see Table S1 in SM). Specifically, in PbClF and PbBrF, holes are approximately 4.6 and 6.8 times heavier than electrons due to the flatter curvature of the valence band(cf. SFig. 1 in SM). Hence, the $\sigma$ of PbClF and PbBrF JLs under electron doping is approximately 6 and 4 times higher, respectively, than that under hole doping. This pronounced asymmetry in $\sigma$ between electron and hole carriers is advantageous for thermoelectric performance as the conduction dominated by a single type of charge carrier enhances efficiency, whereas mixed conduction by both electrons and holes can lead to cancellation of generated Seebeck voltage, thereby reducing the overall thermoelectric efficiency of the material~\cite{snyder08_natmat}. In contrast, PbIF exhibits a parabolic curvature in the valence band, causing the holes to be only about 1.6 times heavier than electrons. Consequently, PbIF shows nearly symmetric electrical conductivity under both electron and hole doping conditions (cf. column (b) in Fig.~\ref{fgr:TEx-FJL}). 

The $\sigma$ of $L$-JLs is approximately 1–3 orders of magnitude higher than that of the $F$-JLs, cf. column (b) in Fig.~\ref{fgr:TEx-FJL} and \ref{fgr:TEx-LJL}. This stems from the bandgaps and optimal electron effective mass due to the curvature of the conduction band (cf. SFig. 1 in SM). The estimated bandgap of $L$-JLs is below $\SI{3}{\electronvolt}$, smaller than those of $F$-JLs, while their valence band profiles remain nearly similar. Consequently, the effective mass of electrons in $L$-JLs is significantly lower, leading to enhanced carrier mobility and thus higher $\sigma$. Notably, PbICl exhibits a particularly low electron effective mass, resulting in $\sigma$ values up to 3 orders of magnitude higher than those in other JLs under electron doping. In PbBrCl, $\sigma$ is predominantly contributed by electrons, with the hole contributed $\sigma$ exceeding the electron contributed $\sigma$ in $F$-JLs. Similarly, in PbIBr, the $\sigma$ by electron and hole are an order of magnitude higher than the electron contributed $\sigma$ of $F$-JLs, indicating excellent transport properties in the $L$-JLs.

% L-JL
To be a candidate material for TE applications, a high $S$ and $\sigma$ are desirable, while the $\kappa_e$ should remain as low as possible to achieve exceptional TE efficiency. Following the trends observed in $\sigma$, $\kappa_e$ reaches a maximum of approximately $\SI{30}{\watt\per\meter\per\kelvin}$ for electron carriers in the $F$-JLs. In electron conduction dominated PbClF and PbBrF JLs, the $\kappa_e$ is notably higher for electrons than for holes, whose $\kappa_e$ values are around $\SI{2}{\watt\per\meter\per\kelvin}$. In contrast, PbIF exhibits nearly similar $\kappa_e$ for both electrons and holes, contributing equally to $\kappa_e$ up to $\sim\SI{12}{\watt\per\meter\per\kelvin}$. Similarly in $L$-JLs the $\kappa_e$ reaches up to $\SI{62}{\watt\per\meter\per\kelvin}$ in PbBrCl, a remarkable $\SI{1600}{\watt\per\meter\per\kelvin}$ in PbICl, and $\SI{40}{\watt\per\meter\per\kelvin}$ in PbIBr by electron carriers. The hole transport in these systems is many orders lower, with a maximum of $\approx\SI{175}{\watt\per\meter\per\kelvin}$ in PbICl and sinking to as low as $\SI{2}{\watt\per\meter\per\kelvin}$ in PbBrCl. Such high $\kappa_e$, particularly in PbICl, can significantly reduce the overall TE efficiency of PbXY JLs. However, the TE efficiency can be balanced by optimizing $\sigma$ and $S$ by improving the carrier concentrations in these layers by either electron or hole doping.

\subsection{Thermoelectric  efficiency}
The power factor $PF$ captures the interplay between the $S$ and $\sigma$ as the temperature and carrier concentration vary. Since $S$ and $\sigma$ maximize at different carrier concentrations, their optimal correlation eventually determines the TE efficiency of the materials. The computed $PF$ are presented in Fig.~\ref{fgr:TEx-FJL} and \ref{fgr:TEx-LJL} (column (d)) for $F$- and $L$-JLs, respectively, for various carrier concentrations and temperatures. In $F$-JLs, $PF$ is predominantly governed by electron doping at low carrier concentrations. In contrast, $L$-JLs exhibit different behavior specific to the layer. The $PF$ increases with electron doping in PbBrCl and PbIBr, while hole doping yields the highest $PF$ in PbIBr due to the enhancement of $S$ driven by both temperature and doping compared to electron doping, cf. Fig.~\ref{fgr:TEx-LJL}a of PbIBr. At higher carrier concentrations, $PF$ declines, reducing the overall TE performance. The combined effect of temperature and carrier concentration leads to notable $PF$ values in PbXY JLs. At \SI{300}{\kelvin}, the maximum $PF$ for PbClF, PbBrF, and PbIF JLs under electron doping reaches values of \SI{3.621e-3}{\watt\per\meter\per\kelvin\squared}, \SI{1.563e-3}{\watt\per\meter\per\kelvin\squared}, and \SI{2.364e-3}{\watt\per\meter\per\kelvin\squared}, respectively. These values are comparable to those of Pb-based chalcogenides such as Pb$_2$SSe~\cite{Mamani2024}, Pb$_2$Sb$_2$S$_2$~\cite{YGan2021}, and (PbS)$_2$~\cite{PJin2022}. In comparison to $F$-JLs, the $L$-JLs show higher performance with $PF$ reaching \SI{13.637e-3}{\watt\per\meter\per\kelvin\squared} in PbBrCl under electron doping and \SI{16.738e-3}{\watt\per\meter\per\kelvin\squared} in PbIBr under hole doping. Notably, the opposite carrier type in these layers still gains $PF$ values of only a few $\SI{}{\milli\watt\per\meter\per\kelvin\squared}$, significantly mitigating the bipolar effect and outperforming many other 2D materials by three orders of magnitude~\cite{Angran2025, CDing2023, MLIU2022, D4TA02974G}. The remarkable performance is observed in PbICl JL, which exhibits a $PF$ of \SI{0.312}{\watt\per\meter\per\kelvin\squared} under hole doping and \SI{0.152}{\watt\per\meter\per\kelvin\squared} under electron doping at \SI{300}{\kelvin}, which is the highest among the studied PbXY JLs.

Finally, the figure of merit $ZT$ is a key indicator for assessing the practical viability of TE materials. It combines the normalized $PF$ by the total thermal conductivity $\kappa$ ($=\kappa_e + \kappa_l$), and scaled by the absolute temperature $T$, cf. Eq.~\ref{eq:ZT}. Variations in $\kappa_e$ due to the effective mass of the carrier modulate the $PF$ and thus assist in distinguishing the predominant charge carriers in each layer. The computed $ZT$ for $F$- and $L$-JLs are shown in Fig.~\ref{fgr:TEx-FJL} and ~\ref{fgr:TEx-LJL} (column (e)), over a range of temperatures and carrier concentrations. Due to the absence of experimental data for the melting temperature for PbXY JLs, the $ZT$ is evaluated up to \SI{1000}{\kelvin} as representative test cases.

In the case of $F$-JLs, only modest $ZT$ of \SI{0.21}{}, \SI{0.14}{}, and \SI{0.07}{} are reached at \SI{300}{\kelvin} for PbClF, PbBrF, and PbIF, respectively. These peaks are attained at carrier concentrations of $\SI{7.23e19}{\per\centi\metre\squared}$ (electron), $\SI{201.11e19}{\per\centi\metre\squared}$ (hole), and $\SI{39.56e19}{\per\centi\metre\squared}$ (electron). Notably, $ZT$ increases with temperature, but decreases as $N$ increases. At \SI{1000}{\kelvin}, the maximum $ZT$ rise to \SI{2.86}{}, \SI{1.38}{}, and \SI{1.50}{} for PbClF, PbBrF, and PbIF, respectively, occurring at carrier concentrations of $\SI{6.73e19}{\per\centi\metre\squared}$ (electron), $\SI{199.74e19}{\per\centi\metre\squared}$ (hole), and $\SI{48.38e19}{\per\centi\metre\squared}$ (electron). These results suggest that $F$-JLs are promising candidates for thermoelectric applications under high-temperature conditions, where they can achieve substantial efficiency.

A distinct TE behavior with a strong preference for a single carrier type is observed in $L$-JLs. In general, hole doping governs the TE efficiency across all $L$-JLs. At \SI{300}{\kelvin}, the maximum $ZT$  of \SI{0.44}{}, \SI{8.94}{}, and \SI{8.61}{} are achieved in PbBrCl, PbICl, and PbIBr, respectively, corresponding to hole concentrations of $\SI{93.90e19}{\per\centi\metre\squared}$, $\SI{10.57e19}{\per\centi\metre\squared}$, and $\SI{6.13e19}{\per\centi\metre\squared}$. With increasing temperature, these $ZT$ significantly improve, reaching upto \SI{5.15}{}, \SI{27.97}{}, and \SI{36.31}{} at \SI{1000}{\kelvin} under the hole concentrations of $\SI{101.30e19}{\per\centi\metre\squared}$, $\SI{3.32e19}{\per\centi\metre\squared}$, and $\SI{2.30e19}{\per\centi\metre\squared}$, respectively. The elevated temperature reduces the hole concentration requirement to attain peak $ZT$ and pins the optimal carrier concentration closer to the Fermi level. As a secondary effect, the electron doping yields substantial $ZT$ of \SI{15.87}{} and \SI{12.75}{} at \SI{1000}{\kelvin} in PbICl and PbIBr, respectively. The high $ZT$ in these two compounds stems from the optimal interplay between $\sigma$ and $\kappa_e$, along with a low $\kappa_l$, particularly in PbICl, which exhibits the lowest $\kappa_l$ among all PbXY JLs (cf. Fig.~\ref{fgr:kl-total}(a)). Considering the combined effects of $\sigma$, $\kappa_e$, and $\kappa_l$, the $L$-JLs, particularly in PbICl and PbIBr, are strong candidates for practical TE applications at room and elevated temperatures.

\section{Conclusions}
In this study, spin-polarized first-principles calculations are carried out to investigate the thermoelectric performance of lead halide JLs. Based on anions, these layers are categorized into two groups to explore the relationship between their structural and electronic properties to probe the curvature of valence and conduction bands. The electron relaxation behavior is analyzed using deformation potential theory, revealing high carrier mobility in PbXY JLs at room temperature. Particularly, the mobility of charge carriers exhibits anisotropy along the armchair and zigzag directions, with holes scattering more rapidly than electrons. A correlation is found between carrier relaxation time, mobility, and electrical conductivity, indicating that the deformation potential approach reliably explores the carrier dynamics. The electronegativity difference between $X$ and $Y$ halides significantly influences the charge transport properties and acts as a driving force for enhanced electrical conductivity. The lattice vibrations are analyzed to estimate the lattice thermal conductivity, and the three-phonon interactions reveal low lattice thermal conductivity, which is governed by acoustic phonon modes. Using the linearized Boltzmann transport equation under the constant relaxation time approximation, we calculated thermoelectric coefficients, showing that electron conductivity dominates over hole conductivity in most PbXY JLs. The Seebeck coefficient increases with temperature and retains almost similar high peak value in all layers. The power factor improves significantly under electron doping, confirming that electrical conductivity is the primary driver of thermoelectric efficiency. When combined with low thermal conductivity, these power factors lead to an excellent figure of merit $ZT$, reaching up to \SI{8.94}{} at room temperature and increasing up to \SI{36.31}{} at \SI{1000}{K}. Overall, our findings emphasize the potential of two-dimensional lead halide JLs as efficient thermoelectric materials and strongly support their application in next-generation thermoelectric devices.
\section*{Author Contributions}
All authors have contributed equally
\section*{Conflicts of interest}
There are no conflicts to declare.

\section*{Acknowledgements}
MV gratefully acknowledges Prof. Dirk C. Meyer for providing an excellent research environment at the Center for Efficient High-Temperature Processes and Materials Conversion (ZeHS), TU Bergakademie Freiberg. MV thanks the Department of Information Services and Computing at Helmholtz-Zentrum Dresden-Rossendorf for access to high-performance computing resources. Furthermore, MV and MZ also acknowledge computing time on the cluster of the Faculty of Mathematics and Computer Science at Technische Universit\"at Bergakademie Freiberg, operated by the University Computing Center (URZ) and funded by the Deutsche Forschungsgemeinschaft under grant number 397252409.

%% REFERENCES %%
\section*{Reference}
\balance
\bibliographystyle{iop}
\bibliography{reference}
\end{document}